\documentstyle[12pt]{article}
\def\barr{\begin{array}}
\def\earr{\end{array}}
\def\beq{\begin{equation}}
\def\eeq{\end{equation}}
\def\bea{\begin{eqnarray}}
\def\eea{\end{eqnarray}}
\def\bmath{\begin{displaymath}}
\def\emath{\end{displaymath}}
\def\bq{\begin{quote}}
\def\eq{\end{quote}}
\def\cA{{\cal A}}

\def\cP{{\cal P}}
\def\cT{{\cal T}}

\def\lh{\lambda_h}
\def\lR{\lambda_R}
\def\PL{\mbox{P}_L}
\def\PR{\mbox{P}_R}
\def\veps{\varepsilon}
\def\apprle{\hspace{-0.1cm}\stackrel{\displaystyle <}{\sim}}

\def\slash#1{\setbox0=\hbox{$#1$}#1\hskip-\wd0\hbox to\wd0{\hss\sl/\/\hss}}

\def\lNi{\lambda_{N_i}}
\def\lNj{\lambda_{N_j}}
\def\lZ{\lambda_Z}
\def\lN1{\lambda_{N_1}}
\def\mpla#1{{\em Mod.\ Phys.\ Lett.\ }{\bf A#1}}

\def\npb#1{{\em Nucl.\ Phys.\ }{\bf B#1}}
\def\plb#1{{\em Phys.\ Lett.\ }{\bf A#1}}
\def\plb#1{{\em Phys.\ Lett.\ }{\bf B#1}}
\def\prl#1{{\em Phys.\ Rev.\ Lett.\ }{\bf #1}}

\def\prd#1{{\em Phys.\ Rev.\ }{\bf D#1}}

\def\prep#1{{\em Phys.\ Rep.\ }{\bf #1}}
\def\ptp#1{{\em Prog.\ Theor.\ Phys.\ }{\bf #1}}

\def\zpc#1{{\em Z.\ Phys.\ }{\bf C#1}}
\catcode`\@=11
\long\def\@makefntext#1{
\protect\noindent \hbox to 3.2pt {\hskip-.9pt
$^{{\ninerm\@thefnmark}}$\hfil}#1\hfill}		%CAN BE USED

\def\@makefnmark{\hbox to 0pt{$^{\@thefnmark}$\hss}}  %ORIGINAL

\def\ps@myheadings{\let\@mkboth\@gobbletwo
\def\@oddhead{\hbox{}
\rightmark\hfil\ninerm\thepage}
\def\@oddfoot{}\def\@evenhead{\ninerm\thepage\hfil
\leftmark\hbox{}}\def\@evenfoot{}
\def\sectionmark##1{}\def\subsectionmark##1{}}

\renewcommand{\thefootnote}{\fnsymbol{footnote}}

\newcounter{sectionc}\newcounter{subsectionc}\newcounter{subsubsectionc}
\renewcommand{\section}[1] {\vspace*{0.6cm}\addtocounter{sectionc}{1}
\setcounter{subsectionc}{0}\setcounter{subsubsectionc}{0}\noindent
	{\normalsize\bf\thesectionc. #1}\par\vspace*{0.4cm}}
\renewcommand{\subsection}[1] {\vspace*{0.6cm}\addtocounter{subsectionc}{1}
	\setcounter{subsubsectionc}{0}\noindent
	{\normalsize\it\thesectionc.\thesubsectionc. #1}\par\vspace*{0.4cm}}
\renewcommand{\subsubsection}[1]
{\vspace*{0.6cm}\addtocounter{subsubsectionc}{1}
	\noindent {\normalsize\rm\thesectionc.\thesubsectionc.\thesubsubsectionc.
	#1}\par\vspace*{0.4cm}}

\newcounter{appendixc}
\newcounter{subappendixc}[appendixc]
\newcounter{subsubappendixc}[subappendixc]

\renewcommand{\appendix}[1] {\vspace*{0.6cm}
        \refstepcounter{appendixc}
        \setcounter{figure}{0}
        \setcounter{table}{0}
        \setcounter{equation}{0}
        \renewcommand{\thefigure}{\Alph{appendixc}.\arabic{figure}}
        \renewcommand{\thetable}{\Alph{appendixc}.\arabic{table}}
        \renewcommand{\theappendixc}{\Alph{appendixc}}
        \renewcommand{\theequation}{\Alph{appendixc}.\arabic{equation}}
        \noindent{\bf Appendix \theappendixc #1}\par\vspace*{0.4cm}}

\def\abstracts#1{{

\centering{\begin{minipage}{12.2truecm}\footnotesize\baselineskip=12pt\noindent
	\centerline{\footnotesize ABSTRACT}\vspace*{0.3cm}
	\parindent=0pt #1
	\end{minipage}}\par}}

\renewenvironment{thebibliography}[1]
	{\begin{list}{\arabic{enumi}.}
	{\usecounter{enumi}\setlength{\parsep}{0pt}
\setlength{\leftmargin 1.25cm}{\rightmargin 0pt}
	 \setlength{\itemsep}{0pt} \settowidth
	{\labelwidth}{#1.}\sloppy}}{\end{list}}

\newcounter{itemlistc}
\newcounter{romanlistc}
\newcounter{alphlistc}
\newcounter{arabiclistc}

\newcommand{\fcaption}[1]{
        \refstepcounter{figure}
        \setbox\@tempboxa = \hbox{\footnotesize Fig.~\thefigure. #1}
        \ifdim \wd\@tempboxa > 6in
           {\begin{center}
        \parbox{6in}{\footnotesize\baselineskip=12pt Fig.~\thefigure. #1}
            \end{center}}
        \else
             {\begin{center}
             {\footnotesize Fig.~\thefigure. #1}
              \end{center}}
        \fi}

\newcommand{\tcaption}[1]{
        \refstepcounter{table}
        \setbox\@tempboxa = \hbox{\footnotesize Table~\thetable. #1}
        \ifdim \wd\@tempboxa > 6in
           {\begin{center}
        \parbox{6in}{\footnotesize\baselineskip=12pt Table~\thetable. #1}
            \end{center}}
        \else
             {\begin{center}
             {\footnotesize Table~\thetable. #1}
              \end{center}}
        \fi}

\def\@citex[#1]#2{\if@filesw\immediate\write\@auxout
	{\string\citation{#2}}\fi
\def\@citea{}\@cite{\@for\@citeb:=#2\do
	{\@citea\def\@citea{,}\@ifundefined
	{b@\@citeb}{{\bf ?}\@warning
	{Citation `\@citeb' on page \thepage \space undefined}}
	{\csname b@\@citeb\endcsname}}}{#1}}

\newif\if@cghi
\def\cite{\@cghitrue\@ifnextchar [{\@tempswatrue
	\@citex}{\@tempswafalse\@citex[]}}
\def\citelow{\@cghifalse\@ifnextchar [{\@tempswatrue
	\@citex}{\@tempswafalse\@citex[]}}
\def\@cite#1#2{{$\null^{#1}$\if@tempswa\typeout
	{IJCGA warning: optional citation argument
	ignored: `#2'} \fi}}

 1
 1
 1

\font\ninerm=cmr9

\begin{document}
\begin{flushright}
RAL-95-025\\
March 1995
\end{flushright}

\centerline{\normalsize\bf SLC/LEP CONSTRAINTS ON UNIFIED MODELS}
\baselineskip=30pt

%\vfill
%\vspace*{0.6cm}
\centerline{\footnotesize APOSTOLOS PILAFTSIS\footnote{Invited
talk given at Ringberg Workshop on ``Perspectives for electroweak
interactions in $e^+e^-$ collisions'', (February 5 -- 8, Munich, 1995).}}
\baselineskip=13pt
\centerline{\footnotesize\it Rutherford Appleton Laboratory, Chilton,
Didcot, Oxon, OX11 0QX, UK}
\centerline{\footnotesize E-mail: pilaftsis@v2.rl.ac.uk}
%\vfill
\vspace*{0.9cm}
\abstracts{
We examine the potential of constraining possible nondecoupling effects of
heavy neutrinos and Higgs bosons at LEP and SLC that may show up in the
nonoblique part of the $Zl_il_j$ couplings.
We analyze this type of  new-physics  interactions  within  the  context  of
low-energy scenarios motivated by unified theories,  such  as  the  Standard
Model (SM) with neutral isosinglets, the left-right symmetric model, and the
minimal supersymmetric SM. Our analysis comprises a complete set of physical
quantities based on the nonobservation of flavour-violating  $Z$-boson decays,
lepton universality in the decays $Z\to l\bar{l}$, and universality of lepton
asymmetries at the $Z$  peak. It is found that these quantities form  a  set
of complementary observables and may hence constrain the parameter  space of
the theories. Non-SM contributions of  new-physics  interactions  to
$R_b=\Gamma(Z\to b\bar{b})/\Gamma(Z\to \mbox{hadrons})$ are briefly discussed
within these models.}

%\vspace*{0.6cm}
\normalsize\baselineskip=15pt
\setcounter{footnote}{0}
\renewcommand{\thefootnote}{\alph{footnote}}
\section{Introduction}
In view of the recent discrepancy of about $2\sigma$ standard
deviations between the leptonic asymmetry $\cA_e$ measured at
the Large Electron Positron Collider (LEP)~\cite{CERN} and the
left-right asymmetry $\cA_{LR}$ at the Stanford Linear Collider
(SLC),\cite{SLD} one may have to
face the fact that the minimal Standard Model (SM) may not be the
underlying theory of nature.~\cite{Ross} If our understanding of
nature is due to some unified theory, such as supersymmetry (SUSY),
grand unified theories (GUTs), superstrings, etc., it is then important
to know the size of new-physics effects expected to come from such theories
at LEP and SLC. Analyzing electroweak oblique parameters has
become a common strategy to test the viability of models beyond
the SM, especially when new physics couples predominantly to $W$ and $Z$
bosons.\cite{STU,DKS} However, one has to explore additional observables
that could be more sensitive to other sectors of the SM.

In Section 2, we will therefore focus our discussion on observables
exhibiting lepton-universality and lepton-flavour violation via the
$Zl_1l_2$ couplings and describe the experimental situation at LEP/SLC.
Then, we will analyze new-physics interactions in the
leptonic sector within the context of low-energy extensions of the
SM that could be motivated by SUSY-GUTs, such as SUSY-$SO(10)$.
Such theories~\cite{SUSYGUT} have received much attention, since the
the electroweak mixing angle, $\sin^2\theta_w (\equiv s^2_w)$,
\newpage
\noindent
predicted is in excellent agreement with its value measured experimentally.
Also, the supersymmetric nature of a SUSY-GUT model prefers
higher unification-point values than usual GUTs, which makes
proton practically ``stable'' with a lifetime of order $10^{36}-10^{38}$
years.
Obviously, the low-energy limit of a SUSY-GUT scenario depends crucially
on the field content and the
details of the breaking mechanism from the unification scale down to
the electroweak one.\cite{Ma} In particular, we shall discuss
the phenomenological implications of three representative extensions
of the SM for the leptonic sector, which could also be the low-energy
limit of certain SUSY-GUTs. Thus, Sections~3, 4, and 5
deal correspondingly with the SM with
left-handed and/or right-handed neutral isosinglets,
the left-right symmetric model (LRSM), and  the minimal SUSY-SM.
In Section~6, we will briefly discuss predictions obtained for
$R_b$ within these models. We draw our conclusions in
Section 7.

\section{The {\boldmath $Zl_1l_2$} vertex}

Here, we  define more precisely the framework of our
calculations. In the limit of vanishing charged lepton masses, the
decay amplitude for $Z\to l_1\bar{l}_2$
can generally be parametrized as
\beq
\cT_l\ =\ \frac{ ig_w}{ 2c_w}\,
\veps^\mu_Z\, \bar{u}_{l_1}\gamma_\mu [g^{l_1l_2}_L\PL\
+\ g^{l_1l_2}_R\PR ] v_{l_2}, \label{Ampl}
\eeq
where $g_w$ is the usual electroweak coupling constant, $\veps_Z^\mu$ is
the $Z$-boson polarization vector, $u\ (v)$ is the Dirac spinor of
the charged lepton $l_1\ (l_2)$,
$\PL(\PR)=(1-(+)\gamma_5)/2$, and $c^2_w=1-s^2_w=M^2_W/M^2_Z$.
In Eq.~(\ref{Ampl}), we have defined
\beq
g^{l_1l_2}_{L,R}=g^l_{L,R}+\delta g^{l_1l_2}_{L,R},\qquad
g^l_L =\sqrt{\rho_l}(1-2\bar{s}^2_w),\qquad g^l_R=-2\sqrt{\rho_l}\bar{s}^2_w,
\eeq
where $\rho_l$, $\bar{s}_w$, $\delta g^{ll}_{L,R}\ (\equiv\delta g_{L,R}^l)$
are obtained beyond the Born approximation and are renormalization-scheme
dependent. In particular, $\rho_l$, $\bar{s}_w$ introduce universal oblique
corrections,\cite{STU,DKS} whereas $\delta g^{l_1l_2}_{L,R}$ represent
flavour-dependent corrections. It is obvious that an
analogous expression is valid for the decay $Z\to b\bar{b}$, as soon as
$b$-quark mass effects can be neglected.

To facilitate our presentation, we reexpress the flavour-dependent
electroweak corrections in terms of the loop functions $\Gamma^L_{l_1l_2}$
and $\Gamma^R_{l_1l_2}$ as follows:
\bmath
\delta g^{l_1l_2}_L\ =\ \frac{\alpha_w}{2\pi} \, \Gamma^L_{l_1l_2}, \qquad
\delta g^{l_1l_2}_R\ =\ \frac{\alpha_w}{2\pi} \, \Gamma^R_{l_1l_2},
\emath
with $\alpha_w=g^2_w/4\pi$.
The nonoblique loop functions $\Gamma^L_{l_1l_2}$
and $\Gamma^R_{l_1l_2}$ depend on whether the underlying theory
is of V--A or V+A nature. Then, the branching ratio for possible
decays of the $Z$ boson into two different charged leptons is given by
\beq
B(Z\to \bar{l}_1 l_2+l_1\bar{l}_2)\ =\ \frac{\alpha_w^3}{48\pi c^2_w}\,
\frac{M_Z}{\Gamma_Z} \Big[ |\Gamma^L_{l_1l_2}|^2+|\Gamma^R_{l_1l_2}|^2\Big].
\label{Blfv}
\eeq
This kind of non-SM decays are constrained by LEP results to be, {\em e.g.},
$B(Z\to e\tau)\apprle 10^{-5}$.~\cite{PDG}

Another observable that has been analyzed recently~\cite{BKPS} is the
universality-breaking parameter $U_{br}^{(l_1l_2)}$.
To leading order of perturbation theory,
$U_{br}^{(l_1l_2)}$ is given by
\bea
U_{br}^{(l_1l_2)} &=& \frac{\Gamma (Z\to l_1\bar{l}_1)\ -\
\Gamma (Z\to l_2\bar{l}_2)}{\Gamma (Z\to l_1\bar{l}_1)\ +\
\Gamma (Z\to l_2\bar{l}_2)}\ - \ U_{br}^{(l_1l_2)}(\mbox{PS})\nonumber\\
&=& \frac{g^l_L (\delta g_L^{l_1} - \delta g^{l_2}_L)\ +\
g^l_R (\delta g^{l_1}_R - \delta g_R^{l_2} )}{g^{l2}_L\ +\ g^{l2}_R}\nonumber\\
&=& U_{br}^{(l_1l_2)}(\mbox{L})\ +\ U_{br}^{(l_1l_2)}(\mbox{R}). \label{Ubr}
\eea
In Eq.~(4), $U_{br}^{(l_1l_2)}(\mbox{PS})$ characterizes known phase-space
corrections coming from the nonzero masses of the charged leptons
$l_1$ and $l_2$ that can always be subtracted, and
\bea
U_{br}^{(l_1l_2)}(\mbox{L})&=&
\frac{g^l_L(\delta g_L^{l_1}- \delta g_L^{l_2})}{(g^{l2}_L+g^{l2}_R)}\ =\
\frac{\alpha_w}{2\pi}\frac{g^l_L}{g^{l2}_L+g^{l2}_R}\, \Re e(\Gamma^L_{l_1l_1}
-\Gamma^L_{l_2l_2}), \\
U_{br}^{(l_1l_2)}(\mbox{R})
&=&\frac{g^l_R(\delta g_R^{l_1}- \delta g_R^{l_2})}{g^{l2}_L+g^{l2}_R}\ =\
\frac{\alpha_w}{2\pi}\frac{g^l_R}{g^{l2}_L+g^{l2}_R}\, \Re e(\Gamma^R_{l_1l_1}
-\Gamma^R_{l_2l_2}).
\eea

Lepton asymmetries --- or equivalently forward-backward asymmetries ---
can also be sensitive to new physics. Here, we will
be interested in experiments at LEP/SLC that measure the observable
\bea
\cA_l \ &=&\ \frac{\Gamma (Z\to l_L \bar{l})\ -\ \Gamma (Z\to l_R \bar{l})}{
\Gamma (Z\to l \bar{l})}\ =\ \frac{g^{ll2}_L\ -\ g^{ll2}_R}{g^{ll2}_L\ +\
g^{ll2}_R} \nonumber\\
 &=& \frac{ g^{l2}_L -g^{l2}_R + 2(g^l_L\delta g^l_L - g_R\delta g^l_R) }{
g^{l2}_L + g^{l2}_R + 2(g^l_L\delta g^l_L + g^l_R\delta g^l_R)}.\label{Al}
\eea
In particular, we use the nonuniversality
parameter of lepton asymmetries~\cite{BP}
\beq
\Delta\cA_{l_1l_2}\ =\ \frac{\cA_{l_1}\ -\ \cA_{l_2}}{\cA_{l_1}\ +\ \cA_{l_2}}
\ =\ \frac{1}{\cA^{(SM)}_l} \Big( U_{br}^{(l_1l_2)}(\mbox{L})\ -\
U_{br}^{(l_1l_2)}(\mbox{R}) \Big)\ -\ U_{br}^{(l_1l_2)}, \label{DA}
\eeq
where $\cA^{(SM)}_l$ may be given by the SM value.
We also emphasize that $U_{br}^{(l_1l_2)}=0$ does
{\em not necessarily} imply $\Delta\cA_{l_1l_2}=0$. For instance,
LRSMs can naturally generate situations, in which
$U_{br}(\mbox{L})\simeq -U_{br}(\mbox{R})$ while $\Delta\cA_{l_1l_2}$
becomes sizeable. Moreover, the physical quantities $U_{br}^{(l_1l_2)}$
and $\Delta\cA_{l_1l_2}$ do not depend explicitly on universal electroweak
oblique parameters.

A recent combined analysis of the LEP/SLC results
regarding lepton universality at the $Z$ peak gives~\cite{CERN,SLD}
\bea
|U_{br}^{(ll')}| & < & 5.\ 10^{-3}\qquad (\mbox{SM}:0),\nonumber\\
\cA_\tau (\cP_\tau) &=& 0.143 \pm 0.010\qquad (\mbox{SM}:0.143),\nonumber\\
\cA_e (\cP_\tau) &=& 0.135 \pm 0.011,\nonumber\\
\cA_{FB}^{(0,l)} &=& 0.0170\pm 0.0016\qquad (\mbox{SM}:0.0153),\nonumber\\
\cA_{LR} (SLC)& =& 0.1637 \pm 0.0075,\label{Exp}
\eea
where theoretical predictions obtained in the SM are quoted in the
parentheses.
Note that $\cA_e$ from $\tau$ polarization is 2$\sigma$ away from
the left-right asymmetry, $\cA_{LR}$, measured at SLC. From
Eq.~(\ref{Exp}), one can deduce $\Delta\cA_{\tau e}\simeq -10\%$
when comparing measurements at LEP and SLC.
However, if one assumes that the measurement of $\cA_{LR}$ is correct, then
one could interpret the experimental sensitivity for $\cA_{LR}$
as a stronger upper bound on new physics with $|\Delta\cA_{\tau e}| < 4\%$.
Furthermore, ongoing SLC experiments are measuring the observable
\beq
A_{LR}^{FB}(f)\ =\ \frac{\Delta\sigma( e^-_Le^+\to f\bar{f})_{FB}
- \Delta\sigma( e^-_Re^+\to f\bar{f})_{FB}}{\Delta\sigma( e^-_Le^+\to
f\bar{f})_{FB} + \Delta\sigma( e^-_Re^+\to f\bar{f})_{FB}}=\frac{3}{4}
\cP_e \cA_f, \label{AFBLR}
\eeq
The forward-backward left-right asymmetry for individual flavours will
be an interesting alternative of testing lepton universality in the
SM in the near future.

\section{The SM with neutral isosinglets}

Here, we will adopt the conventions and the model of Ref.~11,
for the charged- and neutral-current interactions. The model extends the
SM by more than one neutral isosinglets, which allows the presence of
large Dirac components in the general Majorana neutrino mass matrix.
The couplings $WlN_i$ and $ZN_iN_j$ to charged leptons $l$ and heavy Majorana
neutrinos $N_i$ are mediated by the mixings $B_{lN_i}$ and $C_{N_iN_j}$,
respectively. For a model with two-right handed neutrinos, for example,
we have~\cite{MPLA,IP}
\beq
B_{lN_1}\ =\ \frac{\rho^{1/4} s^{\nu_l}_L}{\sqrt{1+\rho^{1/2}}}\ , \qquad
B_{lN_2}\ =\ \frac{i s^{\nu_l}_L}{\sqrt{1+\rho^{1/2}}}\ ,
\eeq %5
where $\rho=m^2_{N_2}/m^2_{N_1}$ is the square of the
mass ratio of the two heavy Majorana neutrinos $N_1$ and $N_2$ present
in such a model. The lepton-flavour mixings $s^{\nu_l}_L$ are defined
as:~\cite{LL} $(s^{\nu_l}_L)^2\equiv \sum_{j=1}^{2} |B_{lN_j}|^2$.
Furthermore, the mixings $C_{N_iN_j}$ can be obtained by
$\sum_{l=1}^{3} B_{lN_i}^{\ast}B_{lN_j}  =  C_{N_iN_j}$.
The mixing angles $(s^{\nu_i}_L)^2$ are directly constrained by
low-energy and other LEP data.\cite{BGKLM}
Although some of the constraints could be model-dependent,
we use the conservative upper limits:~\cite{BGKLM}
$(s^{\nu_e}_L)^2$, $(s^{\nu_\mu}_L)^2 < 0.01$, and
$(s^{\nu_\tau}_L)^2 < 0.06$.

Flavour-changing neutral current decays (FCNC) of the
$Z$ boson into two different charged leptons
were found to receive enhancements due to heavy-neutrino
nondecoupling effects.\cite{KPS,MPLA,IP} \footnote{
In general {\em three-generation} Majorana-neutrino mass models,
nondecoupling effects of heavy neutrinos due to large Dirac components,
which result obviously from the spontaneous break-down of the $SU(2)_L$
gauge symmetry, have originally been discussed by the author in relation
with FCNC Higgs boson decays, $H\to l\bar{l}'$.\cite{PLB}}

To leading order of
heavy neutrino masses, the branching ratio of this kind
of decays is given by
\beq
B(Z\to e^-\tau^+ + e^+\tau^-)\ =\ \frac{a^3_w}{768\pi^2c^3_w}\,
\frac{M_W}{\Gamma_Z}\, \frac{m^4_N}{M^4_W}\, (s^{\nu_e}_L)^2
(s^{\nu_\tau}_L)^2\, \Big[ \sum\limits_i (s^{\nu_i}_L)^2\Big]^2,
\eeq
where $\Gamma_Z$ is the total width of the $Z$ boson. An optimistic
theoretical prediction of these decay modes gives
$B(Z\to e\tau)<10^{-6}$, which should be compared with
the present experimental sensitivity of order $10^{-5}$.
On the other hand, taking
$\lN1 = m^2_{N_1}/M^2_W \gg 1$ and
$\rho=m^2_{N_2}/m^2_{N_1} \ge 1$ into account,\cite{BKPS}
the universality-breaking parameter $U_{br}$ can compactly be given by
\bea
U_{br}^{(ll')}\ =\ U_{br}^{(ll')}(\mbox{L}) &=& -\frac{\alpha_w}{8\pi}\,
\frac{g_L}{g^2_L+g^2_R}\, \Big( (s_L^{\nu_l})^2-(s_L^{\nu_{l'}})^2 \Big)
\Bigg[ 3\ln\lN1
               \nonumber\\
   &+&\sum_{i=1}^{n_G}\ (s_L^{\nu_i})^2\;
                  \frac{\lN1 }{(1+\rho^{\frac{1}{2}})^2}\Bigg( 3\rho
                   +\frac{\rho-4\rho^{\frac{3}{2}}+\rho^2}
                         {2(1-\rho)}\ln\rho\Bigg)\Bigg]. \label{UbrSM}
\eea %6

Another attractive low-energy scenario is an extension of the SM
inspired by certain GUTs~\cite{WW} and superstring theories,\cite{EW}
in which left-handed neutral singlets in addition to the right-handed
neutrinos are present.
In this scenario, the light neutrinos are strictly massless
to all orders of perturbation theory,\cite{WW} when $\Delta L=2$
operators are absent from the Yukawa sector.
The minimal case with one left-handed and one right-handed chiral singlets
can effectively be recovered by the SM with two right-handed
neutrinos when taking the degenerate mass limit for the two heavy Majorana
neutrinos in Eq.~(\ref{UbrSM}).
In Table~\ref{tab:tab1},
we present numerical results for both scenarios discussed
above by assuming $m_{N_1}\simeq m_{N_2} = m_N$. The present experimental
upper bound on $U_{br}^{(ll')}$ is $|U_{br}^{(ll')}| < 5.\ 10^{-3}$,~\cite{PDG}
which automatically sets an upper limit
%%%%Table 1
\begin{table}[h]
\tcaption{$U^{\tau e}_{br}$(L), $U^{\tau e}_{br}$(R), and $\Delta\cA_{\tau e}$
in models discussed in Sects.~3 (i), 4 (ii), and 5 (iii).
}\label{tab:tab1}
\small
\begin{tabular}{||rr||r|r||l||}
\hline
 Gauge&Models\hspace{0.5cm}& $U^{\tau e}_{br}(\mbox{L})\ $ &
$U_{br}^{\tau e}(\mbox{R})\  $ &
$\Delta\cA_{\tau e}$ \\
\hline\hline
(i) ~~$(s^{\nu_\tau}_L)^2$ & $m_N\ [$TeV$]$ & &  & \\
      0.035 & 4.0 & $-4.0\ 10^{-3}$ & 0 &$-2.4\ 10^{-2}$ \\
      0.020 & 4.0 & $-2.0\ 10^{-3}$ & 0 &$-1.2\ 10^{-2}$ \\
\hline
(ii) $M_R\ [$TeV$]$ & $M_h\ [$TeV$]$ & & & \\
0.4 & 5    & $-1.\ 10^{-2}$&$7.7\ 10^{-3}$&$-0.13$ \\
0.4 & 50   & $-1.\ 10^{-2}$&$1.5\ 10^{-2}$&$-0.18$ \\
1.0 & 5    & $-1.\ 10^{-2}$&$1.2\ 10^{-3}$&$-0.16$ \\
1.0 & 100  & $-1.\ 10^{-2}$&$3.2\ 10^{-3}$&$-0.10$ \\
\hline
(iii) ~$\theta_L=0$ & $m_{\tilde{l}}\ [$GeV$]$ & & & \\
$\theta_R=\frac{\pi}{2}$ &
              45    & $1.1\ 10^{-4}$ & $-6.4\ 10^{-5}$ &
                                       $1.2\ 10^{-3}$ \\
$m_{\tilde{l}_L}=m_{\tilde{l}_R}$&
              60    & $5.5\ 10^{-5}$ & $-3.1\ 10^{-5}$ &
                                       $6.1\ 10^{-4}$ \\
$=m_{\tilde{l}}$    & 100    & $2.0\ 10^{-5}$ & $-1.1\ 10^{-5}$ &
                                       $2.2\ 10^{-4}$ \\
\hline
\end{tabular}
\end{table}
on $|\Delta\cA_{ll'}| \apprle 3\%$ since $U_{br}^{ll'}(\mbox{R})=0$.

\begin{figure}
\vspace*{13pt}
\leftline{\hfill\vbox{\hrule width 6cm height0.001pt}\hfill}
\vspace*{3.5truein}		%ORIGINAL SIZE=1.6TRUEIN x 100% - 0.2TRUEIN
\leftline{\hfill\vbox{\hrule width 6cm height0.001pt}\hfill}
\fcaption{Feynman diagrams inducing a right-handed non-universal $Zl\bar{l}$
coupling in the LRSM.}
\label{fig:fig1}
\end{figure}

\section{The LRSM}

This model is based on the gauge group $SU(2)_L\times SU(2)_R\times
U(1)_{B-L}$. We have worked out the realistic
case (d) in Ref.~20, in which the vacuum expectation values of
the left-handed Higgs triplet $\Delta_L$ and that of $\phi^0_2$
in the Higgs bi-doublet vanish. In the LRSM, FCNC $Z$ boson decays into two
different charged leptons are calculated recently~\cite{LRmodel}
and found to be of comparable order with those of the SM with neutral
isosinglets. In addition,  LRSMs can naturally give rise to both a
left-handed and a right-handed non-universal $Zl\bar{l}$
coupling.\cite{BP,LRmodel}
The expression for $U_{br}^{(ll')}(\mbox{L})$ equals the
one given in Eq.~(6), while
$U_{br}^{(ll')}(\mbox{R})$ can be obtained by calculating
the Feynman graphs shown in Fig.~\ref{fig:fig1}. In the limit
where the charged gauge bosons $W^\pm_R$
and the charged Higgs bosons $h^\pm$ are
much heavier than the $Z$ boson, the dominant nondecoupling
heavy neutrino and Higgs-scalar contributions
to $U_{br}^{(ll')}(\mbox{R})$ can be cast into the form~\cite{BP}
\bea
U_{br}^{(ll')}(\mbox{R}) &=& \frac{\alpha_w}{8\pi}\,
\frac{g_R}{g^2_L+g^2_R}\, \Big( B^R_{lN_i}B^{R\ast}_{lN_j}-
B^R_{l'N_i}B^{R\ast}_{l'N_j} \Big) \sqrt{\lNi\lNj}\nonumber\\
&&\times\ \Big[\, \delta_{ij} F_1\ +\ C^L_{N_iN_j}F_2\ +\
C^{L\ast}_{N_iN_j}F_3\, \Big], \label{UbrR}
\eea %8
where $F_1$, $F_2$, and $F_3$ are form factors given by
\bea
F_1 &=& 4s^2_\beta [ C_0(\lR,\lR,\lNi)\ -\ C_0(\lR,\lh,\lNi) ], \\
F_2 &=& 2[ C_0(\lR,\lh,\lNi)\ +\ C_0(\lR,\lh,\lNj )\ -\ C_0(\lNi,\lNj,\lR)]
\nonumber\\
&&+\ s^2_\beta [ C_{24}(\lh,\lh,\lNi )\ +\ C_{24}(\lh,\lh,\lNj )\ -\
C_{24}(\lR,\lh,\lNi )\nonumber\\
&& -\ C_{24}(\lR,\lh,\lNj)\ +\ C_{24}(\lNi,\lNj,\lR)\ -\
C_{24}(\lNi,\lNj,\lh ) ]\nonumber\\
&&+\ \frac{s^2_\beta}{c^2_\beta}
\, [ C_{24}(\lNi,\lNj,\lh )\ -\ C_{24}( 0, \lNj,\lh )\ -\
C_{24}(\lNi,0,\lh )\nonumber\\
&&+\ C_{24}(0,0,\lh ) ], \\
F_3 &=& -\, \frac{2}{\sqrt{\lNi\lNj}}\, [ C_{24}(\lNi,\lNj,\lR )\ -\
C_{24}(0,\lNj,\lR )\ -\ C_{24}( \lNi,0,\lR )\nonumber\\
&& +\ C_{24}(0,0,\lR )] \ +\
s^2_\beta\, \sqrt{\lNi\lNj}\, C_0(\lNi,\lNj,\lR ),
\eea %9,10,11
with $\lR=1/s^2_\beta=M^2_R/M^2_W$ and $\lh = M^2_h/M^2_W$.
In addition, the first three arguments of the Passarino-Veltman
loop functions, $C_0$ and $C_{24}$, are taken to be zero.
In Eq.~(\ref{UbrR}), $B^R$ and $C^L$ (assuming no left-right mixing)
are mixing matrices parametrizing the couplings $W_RlN$ and $ZNN$,
respectively.
The value of $U_{br}$(R) depends on many kinematic variables, {\em i.e.},
the masses of heavy neutrinos (in our estimates we use $m_N=4$~TeV),
the $W_R$-boson mass ($M_R$), and the charged Higgs
mass ($M_h$).\cite{LRmodel}
Quantum effects of the remaining Higgs scalars are found
to be rather small\cite{LRmodel} --- see also discussion in Section 6.
In our numerical estimates, we use the
typical values $(s^{\nu_\tau}_L)^2 =0.05$ and $(s^{\nu_e}_L)^2 = 0.01$.
{}From Table~\ref{tab:tab1}, one can remark the complementary r\^ole that
$U_{br}$ and $\Delta A$ play to constrain or establish new-physics
effects. Even if $U_{br}^{(\tau e)}$ could
be unobservably small of order $10^{-3}$ for some range of kinematic
variables, $\Delta\cA_{\tau e}$ can be as large as $10\%$ and hence capable
of further constraining the parameter space of the model.

\begin{figure}
\vspace*{13pt}
\leftline{\hfill\vbox{\hrule width 6cm height0.001pt}\hfill}
\vspace*{1.4truein}		%ORIGINAL SIZE=1.6TRUEIN x 100% - 0.2TRUEIN
\leftline{\hfill\vbox{\hrule width 6cm height0.001pt}\hfill}
\fcaption{Graphs contributing to a nonoblique $Zl\bar{l}$ coupling in
the SUSY-SM.}
\label{fig:fig2}
\end{figure}

\section{The minimal SUSY model}

In this model,\cite{HK} the FCNC decay of the $Z$ boson was estimated to be
rather small, having $B(Z\to l_1\bar{l}_2)<1.\ 10^{-8}$.\cite{LFCNC}
In addition, the SUSY-SM can generate nonvanishing values for
$U_{br}^{(ll')}(\mbox{L})$ and $U_{br}^{(ll')}(\mbox{R})$.
These observables can be induced by left-handed and
right-handed scalar leptons (denoted as $\tilde{l}_L$, $\tilde{l}_R$)
as well as scalar neutrinos. A non-zero non-universal
$Zl\bar{l}$ coupling can be produced if
two non-degenerate left-handed or right-handed scalar leptons,
say $\tilde{l}$ and $\tilde{l}'$, are present.
To get a feeling about the size of the effects expected in
this model, we will consider the SUSY limit
of the gaugino sector, where only explicit SUSY-breaking
scalar-lepton mass terms are taken into account.
Then, only two neutralinos, the photino $\tilde{\gamma}$ and the ``ziggsino''
$\tilde{\zeta}$ with mass $m_{\tilde{\zeta}}=M_Z$, will contribute
as shown in Fig.~\ref{fig:fig2}. ``Ziggsino'' is a Dirac fermion composed
from degenerate Majorana states of a zino $\tilde{z}$ (the SUSY partner
of the $Z$ boson) and one of
the higgsino fields. For our illustrations, we will further
assume that only one scalar lepton $\tilde{l}$ is relatively light
whereas the others,
{\em e.g.}~$\tilde{l}'$, are much heavier than $M_Z$. Due to
the decoupling behaviour of softly broken SUSY theories,
we can neglect quantum effects of $\tilde{l}'$.
It is then straightforward to obtain~\cite{BP}
\bea
U_{br}^{(ll')}(\mbox{L}) &=& -\; \frac{\alpha_w}{8\pi}\,
\frac{g_L^4\cos 2\theta_L}{g^2_L+g^2_R}\ \Bigg[ \frac{g^2_R}{g^2_L}
\int_0^1\int_0^1 dxdy\; y\; \ln\left( 1-
\frac{\lZ}{\lambda_{\tilde{l}_L}}yx(1-x)\right)\nonumber\\
&&+\ \lZ\int_0^1\int_0^1 dx dy\; y\; \ln\left(
\frac{\lambda_{\tilde{l}_L}y + \lZ [1-y-y^2x(1-x)]}{\lambda_{\tilde{l}_L}y
+ \lZ (1-y)} \right)\Bigg], \\
U_{br}^{(ll')}(\mbox{R}) &=& -\; \frac{\alpha_w}{8\pi}\,
\frac{g_R^4\cos 2\theta_R}{g^2_L+g^2_R}\ \Bigg[
\int_0^1\int_0^1 dxdy\; y\; \ln\left( 1-
\frac{\lZ}{\lambda_{\tilde{l}_R}}yx(1-x)\right)\nonumber\\
&&+\ \lZ\int_0^1\int_0^1 dx dy\; y\; \ln\left(
\frac{\lambda_{\tilde{l}_R}y + \lZ [1-y-y^2x(1-x)]}{\lambda_{\tilde{l}_R}y
+ \lZ (1-y)} \right)\Bigg],
\eea %12,13
where $\lZ=M^2_Z/M^2_W$, $\lambda_{\tilde{l}_L}=m^2_{\tilde{l}_L}/M^2_W$,
$\lambda_{\tilde{l}_R}=m^2_{\tilde{l}_R}/M^2_W$, and $\theta_L$
($\theta_R$) is a mixing angle between the two left-handed
(right-handed) scalar leptons $\tilde{l}_L$ ($\tilde{l}_R$) and
$\tilde{l}'_L$ ($\tilde{l}'_R$). From Table~\ref{tab:tab1}, we see that
the universality-violating observables $U_{br}$ and $\Delta\cA$ turn out
to be no much bigger than $10^{-3}$. Nevertheless, in other
SUSY extensions, the situation may be different.
For instance, in SUSY models with right-handed neutrinos,
enhancements coming from the SUSY Yukawa sector are expected to enter via
the coupling of the charged higgsinos to leptons and scalar neutrinos.
In such scenarios, $\Delta\cA_{\tau e}$ could then reach an experimentally
accessible level $\sim 10^{-2}$.

\section{The observable $R_b$}

Another observable which will still be of interest is
\beq
R_b \ =\ 0.2202 \pm 0.0020\qquad (\mbox{SM}:0.2158). \label{Rb}
\eeq
Assuming that the LEP measurement is correct,
$R_b$ turns out to be about $2\sigma$
off from the theoretical prediction of the minimal SM.
New physics contributions to $R_b$ can be conveniently calculated
through~\cite{BDV}
\beq
R_b\ =\ 0.22\Big[ 1+0.78\nabla_b^{(SM)}(m_t)-0.06\Delta\rho^{(SM)}(m_t)\Big],
\eeq
where $\nabla_b^{(SM)}(m_t)$ and $\Delta\rho^{(SM)}(m_t)$ contains the
$m_t$-dependent parts of the vertex and oblique corrections, respectively.
Practically, only $\nabla_b^{(SM)}(m_t)$ gives significant negative
contributions to $R_b$, which behave, in the large top-mass limit,
as~\cite{BPS}
\beq
\nabla_b^{(SM)}(m_t)\ \simeq\ -\frac{20\alpha_w s^2_w}{13\pi}\,
\frac{m^2_t}{M^2_Z}.
\eeq
If there are new physics effects contributing to $\nabla_b^{(SM)}(m_t)$,
these can be calculated by
\beq
\nabla_b^{(new)}(m_t)\ =\ \frac{\alpha_w}{2\pi}\,
\frac{g^b_L\Re e(\Gamma^L_{bb}(m_t)-\Gamma^L_{bb}(0))
+g^b_R\Re e(\Gamma^R_{bb}(m_t)-\Gamma^R_{bb}(0))}{g^{b2}_L\ +\ g^{b2}_R}\, ,
\label{Rbnew}
\eeq
where $g^b_L=1-2s_w^2/3$ and $g^b_R=-2s^2_w/3$.

In the following, we will try to address the question whether
there exist possibilities of producing positive contributions
to $R_b$ within the SUSY-SM and LRSM. As has already been noticed
in Section 3, only positive contributions to $R_b$ are of
potential interest, which will help to achieve a better agreement between
theoretical prediction and the experimental value of $R_b$.

In the SUSY-SM, $R_b$ can in principle receive positive contributions
from the large Yukawa coupling of the charged higgsino to the
scalar top quark and $b$ quark.\cite{RbSUSY} Also, $R_b$ can get enhanced
from large $\tan\beta$ scenarios. However, considering a number
of constraints originating from $B(b\to s\gamma)$,\cite{Ali} relic abundances
of the lightest SUSY particle,\cite{KKRW} the net SUSY effect on
$R_b$ is considerably reduced and $R_b$ is found to be 0.2166,
which is about 1.5$\sigma$ below the experimental value given in
Eq.~(\ref{Rb}).\cite{WKK}

In LRSM, we first consider the Feynman graphs of Figs.~\ref{fig:fig1}(m) and
\ref{fig:fig1}(n), where the external
leptons are replaced by $b$-quarks and virtual down-type quarks are running
in the place of charged leptons. The interaction
of the FCNC scalars $\phi^{0r}_2$ and $\phi^{0i}_2$
with the $d,\ s,\ b$ quarks is enhanced,
since the corresponding couplings are proportional to the top-quark mass.
In fact, the FCNC scalars
generate effective $Zb\bar{b}$ couplings of both V--A and V+A nature.
In the limit $M_{\phi^{0r}_2},\ M_{\phi^{0i}_2}\gg M_Z$,
the effective $Zb\bar{b}$ couplings take the simple form
\bea
\Re e(\Gamma^R_{bb}(m_t)-\Gamma^R_{bb}(0)) &=&  \frac{1}{8} |V^R_{tb}|^2
\frac{m^2_t}{M^2_W}\, \left( \,
\frac{\lambda_S +\lambda_I}{2(\lambda_S-\lambda_I)}\,
\mbox{ln}\frac{\lambda_S}{\lambda_I}\, -\, 1\, \right), \label{RbL}\\
\Re e(\Gamma^L_{bb}(m_t)-\Gamma^L_{bb}(0)) &=& - \frac{1}{8} |V^L_{tb}|^2
\frac{m^2_t}{M^2_W}\, \left( \, \frac{\lambda_S +\lambda_I}{2(\lambda_S
-\lambda_I)}\, \mbox{ln}\frac{\lambda_S}{\lambda_I}\, -\, 1\, \right),
\label{RbR}
\eea
where $\lambda_S=M^2_{\phi^{0r}_2}/M^2_W$ and
$\lambda_I=M^2_{\phi^{0i}_2}/M^2_W$.
The analytic function in the parentheses of the r.h.s.\ of
Eqs.~(\ref{RbL}) and (\ref{RbR}) is always positive and equals zero
when the two scalars $\phi^{0r}_2,\ \phi^{0i}_2$ are degenerate.
Substituting Eqs.~(\ref{RbL}) and (\ref{RbR}) into Eq.~(\ref{Rbnew}),
one easily finds that the SM value of $R_b$ is further decreased.
This may lead to the mass restriction
\beq
M_{\phi^{0r}_2}\ \simeq\  M_{\phi^{0i}_2}. \label{phi}
\eeq
The mass relation~(\ref{phi}) has been used in our numerical
estimates. Other quantum corrections that could help to produce
positive contributions to $R_b$ are due to diagrams similar to
Figs.~\ref{fig:fig1}(h)
and \ref{fig:fig1}(d). Indeed, an analogous calculation gives
\beq
\Re e(\Gamma^R_{bb}(m_t)-\Gamma^R_{bb}(0)) \ =\ -\frac{1}{4} |V^R_{tb}|^2
\frac{m^2_t}{M^2_W}\, s^2_\beta c^2_\beta\,
\left(\, \frac{\lambda_h +\lambda_R}{2(\lambda_h
-\lambda_R)}\, \mbox{ln}\frac{\lambda_h}{\lambda_R}\, -\, 1\, \right).
\label{Rbh+}
\eeq
However, the l.h.s.\ of Eq.~(\ref{Rbh+}) is proportional to $s^2_\beta=
M^2_W/M^2_R$ yielding a rather small positive effect.
The latter simply demonstrates the difficulty of radiatively inducing
positive contributions to $R_b$ within the LRSM.

\section{Conclusions}

We have found that
lepton-flavour-violating $Z$-boson decays, lepton universality
in the decays $Z\to l\bar{l}$, and universality of leptonic
asymmetries form a set of complementary observables, so as
to impose interesting limitations on model-building in the
leptonic  sector. To precisely demonstrate this, we have analyzed
conceivable low-energy scenarios of unified theories, such as
the SM with neutral isosinglets, the left-right symmetric model,
and the minimal SUSY model. In particular, LRSMs can induce sizeable
values for $\Delta\cA_{\tau e}$ at the experimental visible level of
$5-10\%$, whereas the observable $U_{br}$ measuring deviations
from universality in the leptonic partial widths of the $Z$ boson may
turn out to be rather small. As can also be seen from Table~\ref{tab:tab1},
the sign of $\Delta\cA_{\tau e}$ could help to discriminate
among the various theoretical scenarios beyond the SM.
Finally, we have seen that appears rather difficult to
obtain positive contributions to $R_b$ in GUT-motivated scenarios.
For example, in LRSMs, $\delta R_b^{new}$ always tends to be negative,
which favours FCNC scalars that are degenerate in mass.
If the LEP measurement is indeed correct, this may point towards
supersymmetric physics of an underlying theory.
\vspace*{0.9cm}

\noindent
{\bf Acknowledgements.} I wish to thank Jose Bernab\'eu for
discussions, comments, and scientific collaboration. I would also
like to thank Cliff Burgess for helpful conversations and
Bernd Kniehl for discussions and kind hospitality at Max-Planck Institute.


\vspace*{0.9cm}
\noindent

{\bf References}
\begin{thebibliography}{9}

\bibitem{CERN} The LEP collaborations, ALEPH, DELPHI, L3, OPAL,
and LEP electroweak working group, CERN preprint (1994), CERN/PPE/94-187,
contributed to the 27th Int. Conference on High Energy Physics, Glasgow,
20--27 July, 1994.

\bibitem{SLD} SLD collaboration, K.~Abe et al., \prl{73} (1994) 25.

\bibitem{Ross} T.G.~Rizzo, \prd{50} (1994) 2256; P.~Bamert and C.P.~Burgess,
McGill preprint (1994), McGill-94/27; F.~Caravaglios and G.G.~Ross, CERN
preprint (1994), CERN-TH.7474/94; J.~Erler, Pennsylvania preprint (1994),
UPR-0633T.

\bibitem{STU}
M.E.~Peskin and T.~Takeuchi, \prl{65} (1990) 964; \prd{46} (1992) 381;
G.~Altarelli and R.~Barbieri, \plb{253} (1991) 161;
G.~Altarelli, R.~Barbieri, and S.~Jadach, \npb{369} (1992) 3;
I.\ Maksymyk, C.P.\ Burgess, and D.\ London, \prd{50} (1994) 529.

\bibitem{DKS} For a gauge invariant formulation of the $S$, $T$, and
$U$ parameters, see, G.\ Degrassi, B.A.\ Kniehl, and A.\ Sirlin,
\prd{48} (1993) R3963.

\bibitem{SUSYGUT} J.~Ellis, S.~Kelley, and D.V.~Nanopoulos,
\plb{260} (1991) 131; P.~Langacker and M.~Luo, \prd{44} (1991) 817;
G.G.~Ross and R.G.~Roberts, \npb{377} (1992) 571.

\bibitem{Ma} For a unifiable supersymmetric left-right symmetric model, see,
E.\ Ma, \plb{344} (1995) 164; \prd{51} (1995) 236.

\bibitem{PDG} Particle Data Group, M.~Aguilar-Benitez et al., Review of
particle
properties, \prd{50} (1994) 1173.

\bibitem{BKPS} J.\ Bernab\'eu, J.G.\ K\"orner, A.\ Pilaftsis, and K.\
Schilcher,
\prl{71} (1993) 2695.

\bibitem{BP} J.\ Bernab\'eu and A.\ Pilaftsis, RAL report (1994), RAL-94-089
and FTUV-94-56 (hep-ph/9502296).

\bibitem{ZPC} A.~Pilaftsis, \zpc{55} (1992) 275.

\bibitem{MPLA} A.\ Pilaftsis, \mpla{9} (1994) 3595.

\bibitem{IP} A.~Ilakovac and A.~Pilaftsis, RAL preprint (1994), RAL/94-032,
{\em Nucl.\ Phys.\ B} (in press) (hep-ph/9403398).

\bibitem{LL} P.~Langacker and D.~London, \prd{38} (1988) 886.

\bibitem{BGKLM} C.P.\ Burgess et al., \prd{49} (1994) 6115.

\bibitem{KPS} J.G.~K\"orner, A.~Pilaftsis, and K.~Schilcher,
\plb{300} (1993) 381.

\bibitem{PLB} A.\ Pilaftsis, \plb{285} (1992) 68.

\bibitem{WW} D.~Wyler and L.~Wolfenstein, \npb{218} (1983) 205.

\bibitem{EW} E.~Witten, \npb{268} (1986) 79; R.N.~Mohapatra
and J.W.F.~Valle, \prd{34} (1986) 1642; S.\ Nandi and U.\ Sarkar,
\prl{56} (1986) 564.

\bibitem{GGMKO} J.F.~Gunion et al., \prd{40} (1989) 1546; see also,
C.S.~Lim and T.~Inami, \ptp{67} (1982) 1569.

\bibitem{LRmodel} A.\ Pilaftsis, RAL report (1995), RAL-95-014
(hep-ph/9502330)

\bibitem{HK} For a review, see, H.\ Haber and G.\ Kane, \prep{117} (1985) 75.

\bibitem{LFCNC}
M.J.S.~Levine, \prd{36} (1987) 1329;
F.~Gabbiani, J.H.~Kim, and A.~Masiero, \plb{214} (1988) 398.

\bibitem{BDV} F.\ Boudjema, A.\ Djouadi, and C.\ Verzegnassi,
\plb{238} (1990) 423.

\bibitem{BPS} A.A.\ Akhudov, D.Y.\ Bardin, and T.\ Rieman, \npb{276}
(1986) 1;
J.\ Bernab\'eu, A.\ Pich, and A.\ Santamaria,
\plb{200} (1988) 569; \npb{363} (1991)  326;
W.~Beenaker, W.\ Hollik, \zpc{40} (1988) 141.

\bibitem{RbSUSY} A.\ Denner, R.\ Guth, W.\ Hollik, and J.\ K\"uhn,
\zpc{51} (1991) 695; A.\ Djouadi {\em et al.}, \npb{349} (1991) 48;
M.\ Boulware and D.\ Finnell, \prd{44} (1991) 2054;
A.\ Djouadi, M.\ Drees, and H.\ K\"onig, \prd{48} (1993) 3081;

\bibitem{Ali} A.\ Ali, talk given at this workshop.

\bibitem{KKRW} G.\ Kane, C.\ Kolda, L.\ Roszkowski, and J.\ Wells,
\prd{49} (1994) 6173.

\bibitem{WKK} J.D.\ Wells, C.\ Kolda, G.L.\ Kane, \plb{338} (1994) 219.

\end{thebibliography}
\end{document}